\documentclass[conference]{IEEEtran}
\IEEEoverridecommandlockouts
\usepackage{cite}
\usepackage{amsmath,amssymb,amsfonts}
\usepackage{algorithmic}
\usepackage{graphicx}
\usepackage{textcomp}
\usepackage{xcolor}
\usepackage{tikz}
\usepackage{amsfonts}
\usepackage{amsmath}
\usepackage{xurl}
\usepackage{caption}  
\usepackage{tabularx}
\usepackage[numbers]{natbib}
\usepackage{bbm}
\DeclareMathOperator{\E}{\mathbb{E}}

\usetikzlibrary{automata,arrows,positioning,calc}
\usetikzlibrary{fit}
\usepackage{pgfplots}
\usepackage{subcaption}
\usepackage[numbers]{natbib}
\usepackage{float}

\def\BibTeX{{\rm B\kern-.05em{\sc i\kern-.025em b}\kern-.08em
    T\kern-.1667em\lower.7ex\hbox{E}\kern-.125emX}}
\begin{document}

\title{Stochastic Analysis of Retention Time of Coupled Memory Topology\\}

\author{\IEEEauthorblockN{Anirudh Bangalore Shankar, Avhishek Chatterjee, Bhaswar Chakrabarti and Anjan Chakravorty}
\IEEEauthorblockA{Indian Institute of Technology Madras, India}}

\maketitle

\begin{abstract}
Recently, it has been experimentally demonstrated that individual memory units coupled in certain topology can provide the intended performance \cite{Borders2019,Kaiser2022,Grimaldi2022}. However, experimental or simulation based evaluation of different coupled memory topologies and materials are costly and time consuming. In this paper, inspired by Glauber dynamics models in non-equilibrium statistical mechanics, we propose a physically accurate generic mathematical framework for analyzing retention times of various coupled memory topologies and materials. We demonstrate efficacy of the proposed framework by deriving closed form expressions for a few popular coupled and uncoupled memory topologies, which match simulations. Our analysis also offers analytical insights helping us estimate the impact of materials and topologies on retention time. 
\end{abstract}


\section{Introduction}
\textcolor{black}{
With the introduction of Large Language Models, Artificial Intelligence (AI) and Machine Learning (ML) are playing increasingly important roles in modern society. On the other hand, the amount of data generated on a daily basis around the world is reaching a level that is unprecedented in human history. As a result of this data-deluge, the storage and bandwidth requirements has been on a steep rise. On the other hand, processing of enormous amounts of data comes with an energy-cost that is a global concern \cite{Pattersonetal2021arxiv}\cite{Sudhakar2023}. Data centers generate enormous amounts of heat and need millions of gallons of water to cool down. These facts reveal the growing concerns about the long-term sustainability and environmental impacts of memory and computing technologies closely associated with AI.
}




\textcolor{black}{
Traditional memory technology based on the principle of charge storage has reached its allowable limit of down-scaling since extremely small amount of stored charge easily leaks away posing a threat of low reliability and noise margin. The scaling of technology also adversely affects two other serious performance metrics, namely, the memory retention time and write-energy per bit. Observing the representative data in \cite{Shimeng2016} on DRAM and (emerging) non-volatile memory (NVM) performances, it is apparent that the retention time has a close relationship with the write energy per bit. Increasing retention time inevitably requires higher write energy per bit. In the applications demanding NVM technology, the retention time is around 10 years, whereas in applications such as p-bit computing \cite{Chowdhury2023}, the required retention time is a few millisecond. The experimental work demonstrated in \cite{Borders2019}\cite{Kaiser2022}\cite{Grimaldi2022} successfully shows the efficacy of stochastic magnetic tunnel junction (sMTJ) devices for p-bit computing. It is expected that new material research will give rise to more such volatile memory technologies allowing a more aggressive scaling leading to high speed and low write-energy per bit performance; however, the possibility of less reliable performance due to low retention time due to their operation at the quantum boundary cannot be ruled out. A volatile memory with retention time lower than the required value will have to undergo frequent data refreshing or rewriting causing a higher overall write energy when compared with other technology having relatively higher retention time. Improving future volatile memory technology in terms of retention time may come at the cost of higher write energy. Therefore, mathematical analysis of memory retention time may be sufficient at present in order to get an idea of the projected write-energy cost. Technology innovation for future volatile memory may be geared towards breaking this trade-off. In that case, a similar mathematical framework to estimate the write-energy per bit will be useful.
}




\textcolor{black}{
One of the possible way of apparently breaking the trade-off between retention time and write energy per bit is to use redundancy although it attempts to defeat the purpose of footprint reduction by increasing the overall use of costly material. Note that the modern drive towards 3-D memory integration may partly alleviate this issue. At this point, therefore, a relevant query is to investigate if there is any innovative way to obtain a high retention time at low redundancy. A possible affirmative answer to this investigation may be obtained by allowing coupling of the redundant memory-units. Coupling across classical bits is possible as experimentally demonstrated by p-bit computing technology using sMTJs \cite{Chowdhury2023}. Unlike the methodology used in \cite{Chowdhury2023}, the coupling of memory units may be obtained by means of magnetic, thermal or by some other innovative effects. For example, age-old techniques of magnetic coupling \cite{Pohm1966}\cite{Bruyere1969}\cite{Bruyere2003} or their variants may be appropriately used in a controlled manner in order to couple different memory units. More recently, use of Joule heating induced thermal effects are being explored to couple different memory elements in order to achieve energy-efficient computing \cite{Kim2023}\cite{Kumar2020}\cite{Schoen2023}. Although in \cite{Chowdhury2023} the coupling aspect of the p-bit topology was used for probabilistic computation, we intend to explore its use for prolonging memory state. 
}



\textcolor{black}{
In order to investigate the effects of introducing redundancy with and without coupling between the memory units on their retention time, we, in this paper, present a statistical mechanics based  mathematical framework that builds on Markov chain.} This physically well motivated mathematical framework offers analytical tractability and reduces time and cost involved in experimental and simulation based evaluations of retention times for different topologies and materials.  Such a framework may allow us in future to accommodate various approaches towards design and analysis of topological error correction for classical information bits. 

\subsection{Our contribution}
Inspired by the statistical mechanics approach of using Glauber dynamics \cite{WilmerLP2009}  for studying non-equilibrium behavior of kinetic ferromagnetic Ising models \cite{Yang1992,Ito1997}, we introduce a Glauber dynamics based mathematical framework for analyzing degradation of coupled memory units in a thermal bath. In the proposed framework a memory unit is modeled as a spin of a kinetic Ising model and coupling between two memory units is modeled as inter-spin  coupling, a.k.a. ferromagnetic coupling in probability and statistical mechanics literature. The Ising spins evolve according to a Glauber dynamics which faithfully captures the underlying non-equilibrium physics and ensures that the steady state behavior matches the Maxwell-Boltzmann statistics. Using this framework accurate retention time of many coupled memory topologies can be obtained reasonably fast  via analytical and numerical techniques and thus saving significant experimental and simulation efforts.



To demonstrate the efficacy of this model, we analyze a few simple cases of memory topologies and obtain closed form expressions for retention time, which accurately match the simulations. These expressions allow us to compare  across different topologies and  model parameters. In general, these comparisons can also be done via extensive simulations. However,  simulations often take long time, whereas analytical expressions produce accurate results almost instantaneously.

\section{Memory Topology and Temporal Evolution} 
The main goal of this section is to develop an accurate and physically interpretable Glauber dynamics based mathematical framework for modeling evolution of the coupled and uncoupled memory topologies  in a thermal bath. Our approach is inspired and guided by works on kinetic Ising model in classical statistical physics \cite{Yang1992,Ito1997}. We start with the case of single memory unit, which we refer to as single dipole or single spin. Later we build towards a general coupled memory topology.
\subsection{Single dipole}
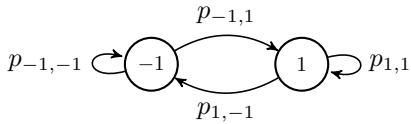
\begin{figure}[H]
    \centering
    \begin{tikzpicture}[->, >=stealth', auto, semithick, node distance=2.5cm]
	\tikzstyle{every state}=[fill=white,draw=black,thick,text=black,scale=0.8]
	\node[state]    (A)  {$-1$};
	\node[state]    (B)[right of=A]   {$1$};
	\path
	(A) edge[loop left]			node{$p_{-1,-1}$}	(A)
        edge[bend left, above]     node{$p_{-1,1}$}  (B)
	(B) edge[loop right]       node{$p_{1,1}$}   (B)
        edge[bend left,below]	node{$p_{1,-1}$}	(A);
	\end{tikzpicture}
 \caption{Markov transition diagram for single dipole.}
 \label{SSDMD}
\end{figure}
The simplest memory topology is a dipole (or spin) representing a single information bit. A dipole orientation $+1$ is for information bit $1$ and $-1$ is for information bit $0$. The dipole is influenced by the heat bath or the environment and its state evolves under the influence of the heat bath. Since almost all  processes in classical statistical physics are memoryless, a continuous time time-homogeneous two state ($-1$ and $1$) Markov chain is an exact model of this evolution \cite{Yang1992,Ito1997,paper2}. 

In physical systems, thermal excitation of a spin happens according to a Poisson process \cite{Yang1992,Ito1997}. The rate of thermal excitation, $\lambda_0$, depends on the the material and the temperature. The Poisson excitation at rate $\lambda_0$ implies that the probability of excitation in a small time interval $(t,t+dt)$ is $\lambda_0 dt$. If and when the dipole is excited, the state of the dipole changes randomly depending on the energy barrier. In general, these random transitions can be modeled using  transition probabilities: $\{p_{i,j}:i,j=\pm 1\}$, where $p_{i,+1}+p_{i,-1}=1$ for any $i=\pm 1$. Thus, the  evolution of the single spin memory is given by the following Markov dynamics:
$$P(\text{state } j \text{ at } t+dt~|~\text{state } i \text{ at } t) = \lambda_0 p_{i,j} dt.$$
The right hand side of the above mathematical equation is simply the product of the probability of excitation in  $(t,t+dt)$ times the probability of $i$ to $j$ transition, since they are independent.


Hence, for the right choice of $p_{i,j}$ the model would be faithful to the  physical system. So, for choosing $p_{i,j}$ we connect  to  the well known equilibrium statistics given by the Maxwell-Boltzmann distribution. As discussed next, $p_{i,j}$ are chosen in such a way that the equilibrium distribution of the above Markov dynamics of the memory is same as the Maxwell-Boltzmann distribution. Such an approach of modeling transient or non-equilibrium behavior by connecting to the Maxwell-Boltzmann distribution has been quite successful for studying transient behavior in statistical mechanics \cite{Yang1992,Ito1997}.

Without loss of generality, throughout the paper we assume that the memory unit is intended to store the information bit $1$ and hence, the corresponding initial state of the dipole is $+1$. In practice, a field may be applied to keep the dipole in its true state $+1$. The potential energy induced by the field on the dipole in state $+1$ and $-1$ be $-H$ and $+H$, respectively. Potential energy can also be induced by retained magnetism or charge of the dipole material. We use $-H$ ($+H$) to model the effective potential energy at dipole state $+1$ ($-1$).

In  classical statistical physics, the probability of crossing a potential barrier $\Delta$ is proportional to $\exp(-\beta \Delta)$ \cite{Yang1992,Ito1997}. Here, $\beta$ represents the effect of temperature of the environment on the barrier; essentially, $1/\beta$ is the temperature times the Boltzmann constant ($kT$). The potential barrier imposed by the memory technology  against change of the state from the initial state ($+1$) to the incorrect state ($-1$) is $2H$. Thus, the probability of transition from $+1$ to $-1$ and $+1$ to $+1$ are proportional to $\exp(-2\beta H)$ and $\exp(\beta 0)(=1)$, respectively. Similar, arguments can be made for $p_{-1,1}$ and $p_{-1,-1}$. Thus by using the fact that $p_{i,-1}+p_{i,1}=1$, expression for $p_{1,-1}$ is obtained as

\begin{equation}
        p_{1,-1} = \frac{e^ {-2\beta H}}{1 + e^ {-2\beta H}},
        p_{-1,1} = \frac{e^ {2\beta H}}{e^ {2\beta H} + 1}.
    \label{eqn2}
    \end{equation}

It can be shown using simple calculations that the two state Markov chain with the above $p_{i,j}$ reaches an equilibrium distribution where the probability of being in a state is proportional to the exponential of the negative of the energy of the state. The well known Maxwell-Boltzmann equilibrium distribution is also of that exact form. This further confirms that the above Markov dynamics with the particular choice of $p_{i,j}$ is indeed the right choice.

\subsection{Coupled $N$-dipoles}
In general, one can construct a memory topology using $N$ dipoles for storing an information bit. An obvious approach to do this would be to store information bit $1$ using $N$ separate dipoles where all are at state $+1$. Each of this dipoles would evolve according to the dynamics of a single dipole, as mentioned above. Note that the stored information bit can be recovered accurately as long as more than $\frac{N}{2}$ dipoles are in $+1$ state. It is intuitive that the retention time of this memory topology would increase with $N$. It would be of  practical interest if we can quantify the retention time as a function with $N$. This would allow us to quantify the cost (due to extra dipoles) versus benefit (increase in retention time) of these topologies. Interestingly, the Markov model for memory evolution discussed in subsection II(a) extends to this case. More importantly, that Markov dynamics extend to a much more general memory topology involving $N$ dipoles. 

Coupling  between two dipoles or spins in a particular way can lower the  energy of the desired state significantly. For example, a positive coupling between the dipoles or spins can lower the potential of all $+1$ state and thus increasing the chance of the memory retaining the information bit $1$. We shall demonstrate that the said Markov dynamics can be extended even to model the coupling between the dipoles.

Consider an undirected graph $G=(V,E)$ with $V$ the set of vertices and $E$ the set of undirected edges. Nodes in $V$ are indexed by integers in $\{1, 2, \ldots, N\}$. Dipoles or spins are placed on the nodes and an edge $(i,j)$ between nodes $i$ and $j$ indicates that the dipoles in those two places are  coupled. 

Associated with each node $i$ is a time-varying  variable $A_i(t)$ taking values $\pm 1$. The initial true state of the system (assuming information bit $1$ is stored) is: $A_i(0)=+1$ for all $i$. The retention time of this memory is the time until which the majority of the variables are $+1$. Now, we venture into the Markov model of stochastic evolution or degradation of the memory.
At any time $t$, the energy of the configuration of the dipoles is 
$$- \sum_i H A_i(t) - \sum_{(i,j) \in E} s_{(i,j)} A_i(t) A_j(t),$$
where $s_{(i,j)}$ is the coupling of the edge $(i,j)$. Here the first term corresponds to the total potential energy of individual dipoles and the second term captures the effect of coupling on the potential energy. Clearly, when coupling is positive, i.e., $s_{(i,j)}>0$, the potential energy is reduced. A positive coupling physically corresponds to a ferro-electric or ferro-magnetic coupling. The memory degradation is influenced by the energy of the configurations and the influence of the environment. Environment excites each dipole independently at a Poisson rate $\lambda_0$. Thus, the total rate at which the dipoles in the memory get excited is $N\lambda_0$. Due to the nature of the Poisson process \cite{WilmerLP2009}, at any time at most one dipole gets excited. On excitation, it randomly changes its state according to the energy barrier, while the states of all the other dipoles remain unchanged. 

Given a configuration $\{A_j(t): j \neq i\}$, the  energy for $i^{th}$ dipole in state $+1$ and $-1$, can be obtained by the above expression. Let us call those energy values $E_i^+$ and $E_i^-$, respectively. Thus, the probability that on excitation $i^{th}$ dipole moves to state $-1$ from $+1$ would be proportional to $\exp(- \beta (E_i^- - E_i^+))$ \cite{Yang1992,Ito1997}. Since the probabilities of transitioning from any state $i$ to $+1$ and $-1$ sum to $1$, the probability of transition from $+1$ to $-1$ is
$\frac{\exp(- 2 \beta (E_i^- - E_i^+))}{1 ~+~\exp(- 2 \beta (E_i^- - E_i^+))}.$
After some algebraic manipulations, we can see that the probability that given $i^{th}$ dipole is excited at time $t$, it moves to state $-1$ from $+1$ is given by
$$\frac{\exp(-\beta \Delta_i(t))}{\exp(\beta \Delta_i(t)) + \exp(- \beta \Delta_i(t))},$$
where $\Delta_i(t)= H + \sum_{j: (i,j) \in E} s_{(i,j)} A_j(t)$.


For any memory topology with $N$ dipoles, the process $\mathbf{A}(t):=\{A_i(t): i \in V\}$, is a continuous time  Markov chain on a $N$-dimensional hypercube $\{\pm 1\}^N$. The retention time of the memory topology is the time by which half the dipoles have flipped from their initial state. In terms of the process $\mathbf{A}(t)$ it is equivalent to $\sum_i A_i(t)<0$.  The nature of the dynamics would depend on the graph $G$ and the coupling values $\{s_{i,j}\}$, and thus, would be the complexity of analyzing the retention time.

\section{Analysis of Different Memory Topologies}

In general, one can always simulate these models for any $N$ and graph $G$. However, the simulations take very large time, especially when the retention time is high. Since in reality we often need to compare between two memory topologies with high retention time, a faster analytical approach would be of practical interest. 

Based on the above Markov chain model, a sequence of equations can be written from which the expected retention time can be obtained. Whenever, there is some symmetry in $G$ and in the choice of $\{s_{i,j}\}$ those equations can also be solved efficiently. Here, we consider only a few special cases with $N=3$ of immediate practical interest. For those configurations analytical  closed form expressions for the retention times are obtained and interesting insights are drawn. The case of large $N$ and $G$ with symmetry will be presented in a future paper.

Mathematically, retention time of a memory can be defined in terms of the process $\mathbf{A}(t)$ discussed above. It is the minimum time to reach a state where the number of dipoles in state $+1$ is less than $\frac{N}{2}$, i.e., $T_{\mbox{ret}} = \min \{t\ge 1: \sum_{i=1}^N A_i(t)<0\}$. This is a random variable and we are interested in obtaining a closed form expression for its expectation $\tau=\mathbf{E}[T_{\mbox{ret}}]$.
\subsection{Single dipole}
In this section, we compute the retention time for a single dipole (i.e.,$N$=1) and analyse the effect of an external magnetic field ($H$) on the same. Here, we use the term 'magnetic' in a generic way to represent any coupling field between two dipoles. The Markov chain for this case is just the state of the dipole and is shown in Fig. \ref{SSDMD}. Initially, at t=0, the dipole is in the +1 state. 

 The retention time here, is the expected time before the dipole enters the -1 state from the initial +1 state. The probability associated with this transition is $p_{1,-1}$. The stochastic process associated with this transition is now a Bernoulli Process with parameter $p_{1,-1}$. The retention time can be thought of as the expected time for the first success in a Bernoulli Process in this case and is given by
\begin{equation}
    \tau = \frac{1}{p_{1,-1}}.
    \label{tauss1}
\end{equation}
In the absence of $H$ field, all the transition probabilities are identical, 
    $p_{1,1} = p_{1,-1} = p_{-1,1} = p_{-1,-1} = 0.50$, 
resulting in a retention time, 
\begin{equation}
         \tau = 2.
         \label{SDH=0}
\end{equation}

In the presence of an external magnetic ($H$) field, the probability of retaining the current state is enhanced if the external field ($H$) parallels the dipole moment. On the other hand, an anti-parallel field will reduce the retention probabilities. Following Fig.~\ref{SSDMD} and the model described in Section II, the transition probabilities are obtained as $1 - p_{-1,-1} = p_{-1,1}$
and $1 - p_{1,-1} = p_{1,1}$.
Using \eqref{eqn2}, one obtains
    \begin{equation}
        p_{1, -1} = \frac{e^ {-\beta H}}{e^ {\beta H} + e^ {-\beta H}}
    \end{equation}
     \begin{equation}
        p_{-1, 1} = \frac{e^ {\beta H}}{e^ {\beta H} + e^ {-\beta H}}
    \end{equation}
leading to the expectation of retention time as
    \begin{equation}
        \tau = \frac{1}{p_{1, -1}} = e^{2 \beta H} + 1.
        \label{SDH}
    \end{equation}

\subsection{Three dipoles}
In this section, we compute the retention time for one bit of memory made from three dipoles. This can be viewed as an extension of the generic N dipole case with N = 3 in this case. The analysis starts with the assumption that all the dipoles are in +1 state. The retention time, by definition, in this system is the time when two dipoles become -1 for the first time. The analysis in this paper is done on three specific graphs: (1) no connection between any of the nodes with each other, (2) triangular graph, and (3) linear graph.
\begin{figure}[H]
\centering
	\begin{tikzpicture}[->, >=stealth', auto, semithick, node distance=2.5cm]
	\tikzstyle{every state}=[fill=white,draw=black,thick,text=black,scale=0.8]
	\node[state]    (A)                     {$0$};
	\node[state]    (B)[right of=A]   {$1$};
	\node[state]    (C)[right of=B]   {$2$};
	\node[state]    (D)[right of=C]   {$3$};
	\path
	(A) edge[loop left]			node{$p_{00}$}	(A)
        edge[bend left, above]     node{$p_{01}$}  (B)
	(B) edge[loop above]       node{$p_{11}$}   (B)
        edge[bend left,below]	node{$p_{10}$}	(A)
	edge[bend left,above]		node{$p_{12}$}	(C)
	(C) edge[loop below]       node{$p_{22}$}   (C)
        edge[bend left,below]	node{$p_{21}$}	(B)
	edge[bend left,above]		node{$p_{23}$}	(D)
	(D) edge[loop right]        node{$p_{33}$}  (D)
        edge[bend left, below]		node{$p_{32}$}	(C);
     \end{tikzpicture}
     \caption{Markov transition diagram for three dipoles.}
     \label{TSDMD}
\end{figure}
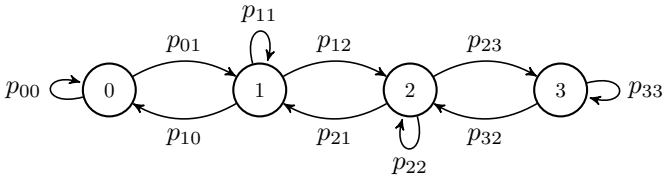
The Markov chain in cases (1) and (2) mentioned above are similar due to inherent symmetry and the respective Markov states are just the number of dipoles in the +1 state. Fig.~\ref{TSDMD} depicts a corresponding Markov transition diagram where the states, $\{0\},\{1\},\{2\}$ or $\{3\}$, represents the number of $+1$ dipoles at any time $t$ for cases (1) and (2). For case (3), the Markov chain is slightly different and is discussed in detail later.
\subsubsection{No connection between any two nodes}
 If $\mathcal{S}$ denotes the number of dipoles in +1 state and $T$ represents the random variable denoting time, using Fig.~\ref{TSDMD} one obtains the following set of linear equations, 
 \begin{equation}
     \E[T|\mathcal{S}=3] = p_{33} \E[T|\mathcal{S}=3] + p_{32} \E[T|\mathcal{S}=2] + 1
     \label{eq1}
 \end{equation}
 \begin{multline}
      \E[T|\mathcal{S}=2] = p_{23} \E[T|\mathcal{S}=3] + p_{22}  \E[T|\mathcal{S}=2] \\+ p_{21} \E[T|\mathcal{S}=1] + 1
      \label{eq2}
 \end{multline}
 \begin{equation}
     \E[T|\mathcal{S}=1] = 0 .
     \label{eq3}
 \end{equation}
Here, $\E[T|\mathcal{S}=s]$ represents the expected "retention time" starting from $s$ number of dipoles being in +1 state. 
The expression $\E[T|\mathcal{S}=3]$ represents the sought-after retention time, $\tau$. \\
In the absence of any $H$ field and assuming no coupling, the necessary transition probabilities appear as 
 \begin{equation}
    p_{33} = p_{32} = 0.50
    \label{eq4}
\end{equation}
and
\begin{equation}
    p_{21} = \frac{1}{3}; p_{22} = \frac{1}{2}; p_{23} = \frac{1}{6}.
    \label{eq5}
\end{equation}
Solving \eqref{eq1}, \eqref{eq2} and \eqref{eq3} using the probabilities given in \eqref{eq4} and \eqref{eq5}, one obtains the retention time as 
    \begin{equation}
       \tau = \E[T|\mathcal{S}=3] = 6.
        \label{eqtau1}
    \end{equation}
The expected retention time per dipole is given by 
\begin{equation}
    \tau^{'} = \frac{\tau}{3} = 2. 
    \label{eqtau2}
\end{equation}
Note, this is thrice the retention time of a single dipole as anticipated.

In the presence of an external magnetic ($H$) field and without any coupling between the dipoles, the probability of each individual dipole going to $-1$ state is the same and is given by $p_{32}$ (see Fig \ref{TSDMD}). Here, the necessary transition probabilities (using Eqn. \eqref{eqn2}) are 
\begin{equation}
    p_{32} = \frac{e^ {-\beta H}}{e^ {\beta H} + e^ {-\beta H}};
    \label{pr32x}
\end{equation}
\begin{equation}
    p_{33} = 1-p_{32} = \frac{e^ {\beta H}}{e^ {\beta H} + e^ {-\beta H}};
    \label{pr1}
\end{equation}
\begin{equation}
    p_{21} = \frac{2}{3} \frac{e^ {-\beta H}}{e^ {\beta H} + e^ {-\beta H}};
    \label{pr2}
\end{equation}
\begin{equation}
    p_{23} = \frac{1}{3} \frac{e^ {\beta H}}{e^ {\beta H} + e^ {-\beta H}};
    \label{pr3}
\end{equation}
\begin{equation}
    p_{22} = 1 - p_{21} - p_{23}.
    \label{pr4}
\end{equation}
The retention time is found by solving equations \eqref{eq1}, \eqref{eq2} and \eqref{eq3} using the probabilities \eqref{pr32x}, \eqref{pr1}, \eqref{pr2}, \eqref{pr3} and \eqref{pr4}. This evaluates to be 
    \begin{equation}
        \tau = \frac{1}{2} (1+e^{2\beta H})^2 + 2(1+e^{2\beta H}).
        \label{Hfe}
    \end{equation}
As in the case of a single dipole, if the magnetic field is in the direction of the dipole moment, the retention time is greatly enhanced. From Eq. \eqref{Hfe}, it is very obvious that for any positive $H$, the retention time is greater than 6 which is the retention time in the case of three isolated dipoles. If $H$ is in the opposite direction the retention time is significantly attenuated. Thus, a careful application of the external field can significantly improve the retention time.

\subsubsection{Triangular graph}
In the presence of any $H$ field in a coupled system of three dipoles arranged in a triangular form, with coupling coefficient $s_f$, the probability of each dipole staying in $+1$ state can be obtained by using \eqref{eqn2} and is given as
    \begin{equation}
        p_{+1} = \frac{e^ {\beta s_f(A[i-1]+A[i+1]) + \beta H}}{e^ {\beta s_f(A[i-1]+A[i+1]) + \beta H} + e^ {- \beta s_f(A[i-1]+A[i+1]) -\beta H}}.
        \label{eqn2ag}
    \end{equation}
The analysis remains exactly the same as in the previous case due to the inherent symmetry of the triangular system. The only change occurs in the transition probabilities. Here, the transition probabilities are obtained from Fig.~\ref{TSDMD} and using \eqref{eqn2ag} and can be expressed as 
    \begin{equation}
         p_{32} = \frac{e^{-2\beta s_f-\beta H}}{e^{2\beta s_f+ \beta H}+e^{-2\beta s_f - \beta H}};
         \label{p1}
    \end{equation}
    \begin{equation}
        p_{21} = \frac{2}{3} \frac{e^{-\beta H}}{e^{\beta H} + e^{-\beta H}};
        \label{p2}
    \end{equation}
    \begin{equation}
        p_{23} = \frac{1}{3} \frac{e^{2\beta s_f+\beta H}}{e^{2\beta s_f+\beta H} + e^{-2 \beta s_f-\beta H}};
        \label{p3}
    \end{equation}
    \begin{equation}
        p_{22} = 1 - p_{23} - p_{21}.
        \label{p4}
    \end{equation}
The other probabilities are not mentioned here since they do not play a role in the calculation of retention time. The resulting retention time can be obtained by solving \eqref{eq1}, \eqref{eq2} and \eqref{eq3} along with the transition probabilities \eqref{p1}, \eqref{p2}, \eqref{p3} and \eqref{p4}.  The closed form expression for the retention time is given as
    \begin{equation}
        \tau = \frac{p_{32}-p_{22}+1}{(1-p_{33})(1-p_{22})-p_{32}p_{23}}.
        \label{ret}
    \end{equation}

\subsubsection{Linear graph}
\begin{figure}[H]
    \centering
    \includegraphics[scale = 0.4]{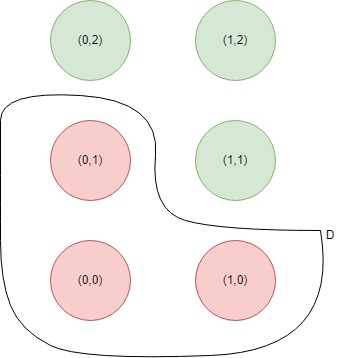}
    \caption{Markov chain evolution for linear graph.}
    \label{fig:efx}
\end{figure}

In the case of a linear graph, the Markov chain is an ordered tuple $(i,j)$, where $i$ and $j$, respectively, represent the numbers of middle ($0$ or $1)$ and boundary dipoles ($0$, $1$ or  $2$) at  $+1$ state. The set $D$ in Fig. \ref{fig:efx} represents the set of states with two or more dipoles in the -1 state. The set $D$ being given by 
\begin{equation*}
    D = \{(0,0), (0,1), (1,0)\}.
\end{equation*}
To evaluate the retention time in this case, we need to solve the following set of linear equations:
\begin{equation}
    \E[T^D_{(1,2)}] = 
    p_{(1,2)}^{(1,2)} \E[T^D_{(1,2)}] +
    p_{(1,2)}^{(1,1)} \E[T_{(1,1)}^D] +
    p_{(1,2)}^{(0,2)} \E[T_{(0,2)}^D] + 1
    \label{l1}
\end{equation}
\begin{equation}
    \E[T^D_{(0,2)}] = 
    p_{(0,2)}^{(1,2)} \E[T^D_{(1,2)}] +
    p_{(0,2)}^{(0,1)} \E[T_{(0,1)}^D] +
    p_{(0,2)}^{(0,2)} \E[T_{(0,2)}^D] + 1
    \label{l2}
\end{equation}
\begin{multline}
    \E[T^D_{(1,1)}] = 
    p_{(1,1)}^{(1,2)} \E[T^D_{(1,2)}] +
    p_{(1,1)}^{(1,1)} \E[T_{(1,1)}^D] +
    p_{(1,1)}^{(1,0)} \E[T_{(1,0)}^D] + \\ p_{(1,1)}^{(0,1)} \E[T_{(0,1)}^D] + 1
    \label{l3}
\end{multline}
\begin{equation}
    \E[T_{(0,1)}^D] =  \E[T_{(1,0)}^D] = \E[T_{(1,1)}^D] = 0
    \label{l4}
\end{equation}
where $p_{(i,j)}^{(x,y)}$ denotes the probability of transitioning from state $(i,j)$ to $(x,y)$ and $\E[T_{(i,j)}^{D}]$ denotes the expected "retention time" to go from state $(i,j)$ to any element of the set $D$.
The probabilities that govern the transition sequence are given as
\begin{equation}
    p_{(1,2)}^{(0,2)} = \frac{1}{3} \frac{e^{-2\beta s_f-\beta H}}{e^{-2\beta s_f-\beta H} + e^{2\beta s_f+\beta H}}
\end{equation}
\begin{equation}
    p_{(1,2)}^{(1,1)} = \frac{2}{3} \frac{e^{-\beta s_f-\beta H}}{e^{-\beta s_f-\beta H} + e^{\beta s_f+\beta H}}
\end{equation}
\begin{equation}
    p_{(1,2)}^{(1,1)} + p_{(1,2)}^{(0,2)} + p_{(1,2)}^{(1,2)} = 1
\end{equation}
\begin{equation}
    p_{(0,2)}^{(0,1)} = \frac{2}{3} \frac{e^{\beta s_f-\beta H}}{e^{\beta s_f-\beta H} + e^{-\beta s_f+\beta H}}
\end{equation}
\begin{equation}
    p_{(0,2)}^{(1,2)} = \frac{1}{3} \frac{e^{2\beta s_f+\beta H}}{e^{-2\beta s_f-\beta H} + e^{2\beta s_f+\beta H}}
\end{equation}
\begin{equation}
    p_{(0,2)}^{(0,1)} + p_{(0,2)}^{(0,2)} + p_{(0,2)}^{(1,2)} = 1
\end{equation}
\begin{equation}
    p_{(1,1)}^{(0,1)} = \frac{1}{3} \frac{e^{-\beta H}}{e^{-\beta H} + e^{\beta H}}
\end{equation}
\begin{equation}
    p_{(1,1)}^{(1,0)} = \frac{1}{3} \frac{e^{-\beta s_f-\beta H}}{e^{-\beta s_f-\beta H} + e^{\beta s_f+\beta H}}
\end{equation}
\begin{equation}
    p_{(1,1)}^{(1,2)} = \frac{1}{3} \frac{e^{\beta s_f+\beta H}}{e^{-\beta s_f-\beta H} + e^{\beta s_f+\beta H}}
\end{equation}
\begin{equation}
    p_{(1,1)}^{(0,1)} + p_{(1,1)}^{(1,0)} + p_{(1,1)}^{(1,2)} +
    p_{(1,1)}^{(1,1)}= 1.
\end{equation}
We need to solve Eqns. \eqref{l1}, \eqref{l2}, \eqref{l3} and \eqref{l4} to get $\E[T_{(1,2)}^D]$ which is the sought-after retention time. 
After solving, we arrive at
\begin{equation}
    \tau = \E[T_{(1,2)}^D] = \frac{\frac{p_{(1,2)}^{(1,1)}}{1-p_{(1,1)}^{(1,1)}}+\frac{p_{(1,2)}^{(0,2)}}{1-p_{(0,2)}^{(0,2)}}+1}{p_{(1,2)}^{(1,1)}+p_{(1,2)}^{(0,2)}-\frac{p_{(1,2)}^{(1,1)}p_{(1,1)}^{(1,2)}}{1-p_{(1,1)}^{(1,1)}}-\frac{p_{(1,2)}^{(0,2)}p_{(0,2)}^{(1,2)}}{1-p_{(0,2)}^{(0,2)}}}.
\end{equation}
    
\section{Comparisons across Memory Topologies}
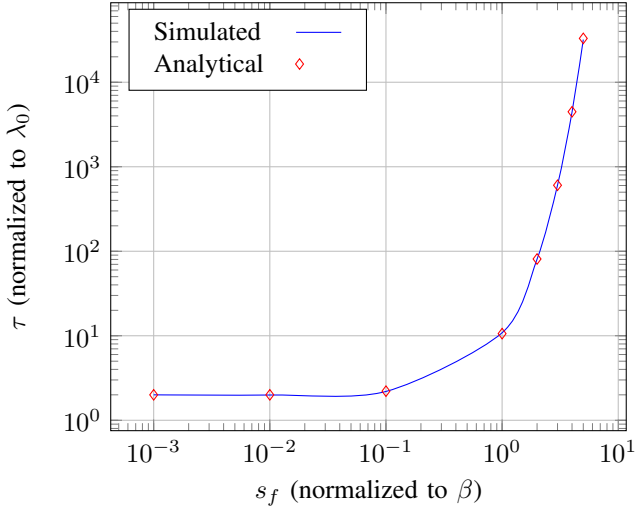
\begin{figure}[H]
    \centering
    \begin{tikzpicture}
            \begin{loglogaxis}
            [name=plot, 
            xlabel={$s_f$ (normalized to $\beta$)}, 
            ylabel={$\tau$ (normalized to $\lambda_0$)},
            legend style={at={(0.5,-0.2)},
            anchor=north},
            grid=major
            ]
            \addplot[blue, smooth] table {./Simulations2.txt}; 
            \label{sinul}
            \addplot[only marks, mark options = {red},  mark = diamond] table{./Analytical2.txt};
            \label{analy}
            \end{loglogaxis}
             \node[anchor=north east ,fill=white,draw=black] (legend) at ($(plot.north)-(0.2 mm, 0.8 mm)$) {\begin{tabular}{l l}
            Simulated & \ref{sinul}  \\ 
            Analytical & \ref{analy} \\
    \end{tabular} };
        \end{tikzpicture}
    \caption{Normalized coupling co-efficient dependent normalized retention time: comparison of results obtained from simulation and analytical expression.}
    \label{fig:anavssim}
\end{figure}

In the previous section, we obtained analytical closed form expressions of the expected retention time, $\tau$, for different memory topologies with $N=3$. In this section, we compare the performance of these topologies using those expressions. First, in Fig.~\ref{fig:anavssim}, we compare the coupling coefficient ($s_f$) dependent retention time obtained from the analytical expression and that from simulation. We considered three dipoles in a linear graph which is the most challenging of all the cases in the previous section. We observe that the analytical expression and the simulations yield almost identical results for different choices of $s_f$.

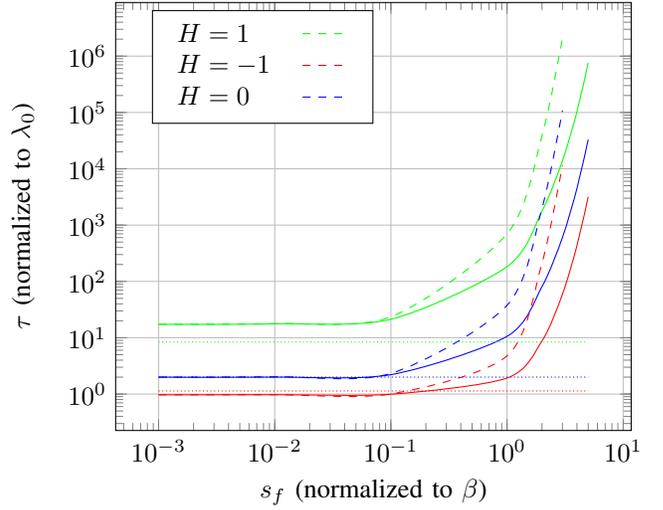
\begin{figure}[H]
    \centering
    \begin{tikzpicture}
            \begin{loglogaxis}
            [name=plot, 
            xlabel={$s_f$ (normalized to $\beta$)}, 
            ylabel={$\tau$ (normalized to $\lambda_0$)},
            legend style={at={(0.5,-0.2)},
            anchor=north},
            grid=major
            ]
            \addplot[red, smooth] table{./Analytical1.txt};
            \label{$B=-1$}
            \addplot[blue, smooth] table{./Analytical2.txt};
            \label{$B=0$}
            \addplot[green, smooth] table{./Analytical3.txt};
            \label{$B=1$}
            \addplot[dashed, red, smooth] table{./CAna1.txt};
            \label{$B=-1$}
            \addplot[dashed, blue, smooth] table{./CAna2.txt};
            \label{$B=0$}
            \addplot[dashed, green, smooth] table{./CAna3.txt};
            \label{$B=1$}
            \addplot[densely dotted, red, smooth] table {Single1.txt};
            \addplot[densely dotted, blue, smooth] table {Single2.txt};
            \addplot[densely dotted, green, smooth] table {Single3.txt};            
            \end{loglogaxis}
             \node[anchor=north east ,fill=white,draw=black] (legend) at ($(plot.north)-(0 mm, 1 mm)$) {\begin{tabular}{l l}
            $H=1$ & \ref{$B=1$}  \\ 
            $H=-1$ & \ref{$B=-1$} \\
            $H=0$ & \ref{$B=0$}  \\  
    \end{tabular} };
        \end{tikzpicture}
    \caption{Normalized coupling coefficient dependent normalized retention time at different external field: comparison among the results obtained for linear array of 3 dipoles (solid), $3$ dipoles in a triangular arrangement (dashed), and single dipole (dotted).}
    \label{fig:N3comparison}
    \end{figure}

In Fig.~\ref{fig:N3comparison}, for $N=3$, we plot the normalized (by $\lambda_0$) retention times of the linear and the triangular topologies against $s_f$ (normalized by $\beta$) for different values of $H$ (normalized by $\beta$). The reason for these normalizations is that both $\lambda_0$ and $\beta$ depend on temperature. The above normalization allows us to compare different topologies independent of the temperature. Note that our model can also analyze the temperature dependence, but we leave that for future work.

Here in Fig.~\ref{fig:N3comparison}, as a baseline, we overlay the plot of the retention time for a single dipole normalized by $\lambda_0$. Note that at small $s_f$ both the triangular and the linear topologies yield similar results as obtained from the uncoupled $N=3$ case. As expected, the curve for $N=3$ uncoupled case is above the single dipole case when $H > 0$. At $H=0$, these two cases match exactly at small $s_f$. However, at $H=-1$, at low $s_f$, the single dipole case has a slightly better retention time. In fact, we can analytically show that when $H$ is a large negative quantity and $s_f=0$, the normalized retention time of the single dipole case is $1$, whereas that of the three dipole case is $\frac{5}{6}$. This implies that if there is an external field which is opposing in nature, having a larger topology is detrimental.

However, $H \ge 0$ is the more likely case representing a helpful external field for a positive retentivity. In that case, we  observe that the retention times are growing exponentially with $s_f$ for the linear and the triangular topology for all the values of $H$. The rate of growth with $s_f$ for the triangular topology is higher.

This seems to imply that the triangular topology is the natural choice for high retention time. However, a proper comparison has to consider that in the triangular topology there is an extra coupling compared to the linear topology. Since an extra coupling would incur extra fabrication and material cost, it is worth quantifying the increase in retention time with that extra coupling. By analyzing the close form expressions, particularly the dominating term, we observe that the retention times of the linear and the triangular topology grow as $\exp(2s_f+2H)$ and $\exp(4 s_f + 2H)$, respectively. This implies that for the same $H$, at any $s_f$, the retention time of triangular case is $\exp(2 s_f)$ times that of the linear case. Thus, the benefit of moving from linear to triangular masks the cost when the material's $s_f$ is high.

The above discussion also brings out the cost benefit trade-off for choosing a costlier material with higher $s_f$. For the same change $\Delta s_f$ in $s_f$, in case of triangular array the relative increase in retention time (ratio of retention times for $s_f$ and $s_f+\Delta s_f$), $\exp(4 \Delta s_f)$, is higher than that of $\exp(2 \Delta s_f)$ obtainable in linear graph. It is interesting that in both cases the relative growth in retention time is exponential in $\Delta s_f$. Therefore, till the point where the material cost is increasing with $s_f$ slower than exponential, it is always better to keep increasing $s_f$.

\section{Conclusion}

In this article, we provided a framework for modeling and analyzing coupled memory topology allowing us to obtain their retention time. The proposed model is physically well motivated from non-equilibrium statistical mechanics and is almost exact. 

We found out that the retention time depends on the material of interest, the coupling coefficient ($s_f$) and the external magnetic ($H$) field. As $s_f$ increases, the retention time rises exponentially. The coefficient in the exponent differs significantly across topology. Addition of an extra coupling to a linear graph increases the retention time by a factor of  $\exp(2s_f)$.  Intuitively, coupling pulls the dipole under consideration to the average state of the surrounded neighbors. Thus, it stabilizes the memory when most of the spins or dipoles are in the correct state. However, it can be detrimental in heavy noise when multiple spins or dipoles are in the wrong state. 

Similarly, the retention time varied exponentially with the applied $H$ field or material retentivity. However, topology do not have an impact on that. If external field is applied in the direction of the dipole moment, the retention time is enhanced significantly. On the other hand, if $H$ is negative, i.e., an external field in anti-parallel direction or the dipole material partially retains a previous opposite state, the retention time is significantly reduced and a larger topology can be detrimental in this case. 

In future, we would like to understand the optimal trade-off between material cost, writing cost and external field using our framework. This would offer insights into the choice of materials and topology for a particular application with a specified range of retention time. 

\bibliographystyle{unsrtnat}
\bibliography{references}
\end{document}